\documentclass[aps, prb, reprint]{revtex4-1}

\usepackage{amsmath}
\usepackage{graphicx}
\usepackage{float}
\usepackage{CJK}
\usepackage{eqnarray}

\begin{document}

\title{Interaction-induced supercurrent in quantum Hall setups}
\date{\today}
\author{Xiao-Li Huang  }
\affiliation{Kavli Institute of NanoScience, Delft University of Technology, Lorentzweg 1, NL 2628 CJ, Delft, The Netherlands}
\author{Yuli V. Nazarov}
\affiliation{Kavli Institute of NanoScience, Delft University of Technology, Lorentzweg 1, NL 2628 CJ, Delft, The Netherlands}
\begin{abstract}
Recently we have proposed an unusual mechanism of superconducting current that is specific for Quantum Hall Edge channels connected to superconducting electrodes. We have shown that the supercurrent can be mediated by a nonlocal electon-electon interaction that provide an opportunity for a long-distance information transfer in the direction opposite to the electron flow. A convenient model for such interaction is that of an external circuit. The consideration has been done for the case of a single channel.

In this work, we extend these results to more sophisticated setups that include the scattering between Quantum Hall channels of opposite direction and multiple superconducting contacts. For a single Quantum Hall constriction, we derive a general and comprehensive relation for the interaction-induced supercurrent in terms of scattering amplitudes and demonstrate the non-local nature of the current by considering its sensitivity to scattering.  We understand the phase dependences of the supercurrents in multi-terminal setups in terms of interference of Andreev reflection processes. For more complex setups encompassing at least two constrictions we find an interplay between non-interacting and interaction-induced currents and contibutions of complex interference processes.
\end{abstract}
\maketitle
\section{Introduction}
Topological edge states, from its first discovery in quantum Hall sample \cite{Klitzing} through the advent of topological insulators\cite{Hasan}, has proved to be both an interesting topic by itself\cite{Huckestein, Murthy, Hansson}, as well as a tool for fundamental physics research and practical area\cite{Weis}. its chirality/helicity and the quantized conductance have both play their own important roles. It is interesting to combine Quantum Hall physics with superconductivity by contacting the edge channels with superconducting electrodes, this being a subject of intensive theoretical and experimental research \cite{Takagaki, Chtchelkatchev, MZ, SL, vOAB, Finkelstein}

The Quantum Hall systems can be devised to arrange the scattering between the edge channels of different chirality, by making corner junctions or constrictions. In Refs.\cite{Ji, Jonckheere, Yang} the authors studied the interferometers made from corner junctions between edge channels for integral and fractional quantum Hall systems. In \cite{Martins}, they have examined tunneling between edge states via an intermediate quantum Hall island. In \cite{Hou}, they have proposed to use a tunnel junction to probe the helicity of edge states. Such Quantum Hall setups can be combined with superconducting electrodes\cite{Finkelstein}. Beenakker \cite{Beenakker} has proposed an experimental setup for probing annihilation probability between Bogoliubov quasiparticles from two superconducting source with  a phase difference between them with the goal to demonstrate their Majorana nature. The Quantum Hall setups may include more than two superconducting electrodes, such multi-terminal superconducting structures are under active theoretical and experimental investigation \cite{Weyl,transconductance,WeylSpin,OrderDisorder,Omega,Padurariu, Amundsen2017}

Recently the authors have addressed the supercurrents in a long chiral edge channel with two superconducting electrodes \cite{HXL&YVN}. While this current vanishes in approximation of non-interacting electrons, we have shown the possibility of an interaction-induced supercurrent. This supercurrent appeared to require a non-local electron-electron interaction and is related to an information flow in the direction opposite to the electron flow that is provided by such interaction. We have considered several model interaction models and formulated an external circuit model that facilitates controllable and efficient non-local interaction.

In this Article, we extend this research to more complex Quantum Hall setups that involve scattering between the edge channels and multiple superconducting electrodes. We compute the interaction-induced supercurrent in these setups. This is important in view of the fact that the setups are easy to employ and flexible to reveal the peculiarities of the effect under consideration. To avoid unnecessary details, we restrict ourselves to the simplest external circuit model of the non-local interaction. 

In all setups, the supercurrent values are of the same order of magnitude as for the single channel case, but do depend on details of potential and Andreev scattering in the structure. Full and general analysis can be performed in a situation of a single constriction where the electron trajectories do not make loops. We specify to several distinct setups, some demonstrating the non-local nature of the interaction-induced supercurrent, some, like the Beenakker setup, not exhibiting any supercurrent at all. In a similar fashion, we analyze the supercurrents for the case of many superconducting terminals connected to a single edge channel. We reveal the relation between the current and a complex amplitude of Andreev scattering that is contributed by a multitude of partial amplitudes corresponding to various sequences of Andreev processes. 
As an example of more complex and potentially interesting situation, we consider a setup comprising two constrictions and two or three superconducting electrodes. The presence of loop trajectories complicates the sequences of Andreev processes and may lead to an interplay of interaction-induced and common proximity supercurrents.


The paper is organized as follows. In Section \ref{sec:IIsupercurrent} we recite the previous results on interaction-induced supercurrent, explain the model and the way to derive the effect microscopically. In Section \ref{sec:single} we apply these concepts to a single junction setup, derive a general formula and specify it to a variety of the situations. In Section \ref{sec:multi} we discuss the supercurrents in multiple superconducting terminals connected to a single edge channel. In Section \ref{sec:complex} we consider a more complex setup comprising two constrictions and three superconducting terminals. We conclude in Section \ref{sec:conclusions}.

\section{Interaction-induced supercurrent in Quantum Hall edge channels}
\label{sec:IIsupercurrent}
\begin{figure}[h]
\includegraphics[width=.5\textwidth]{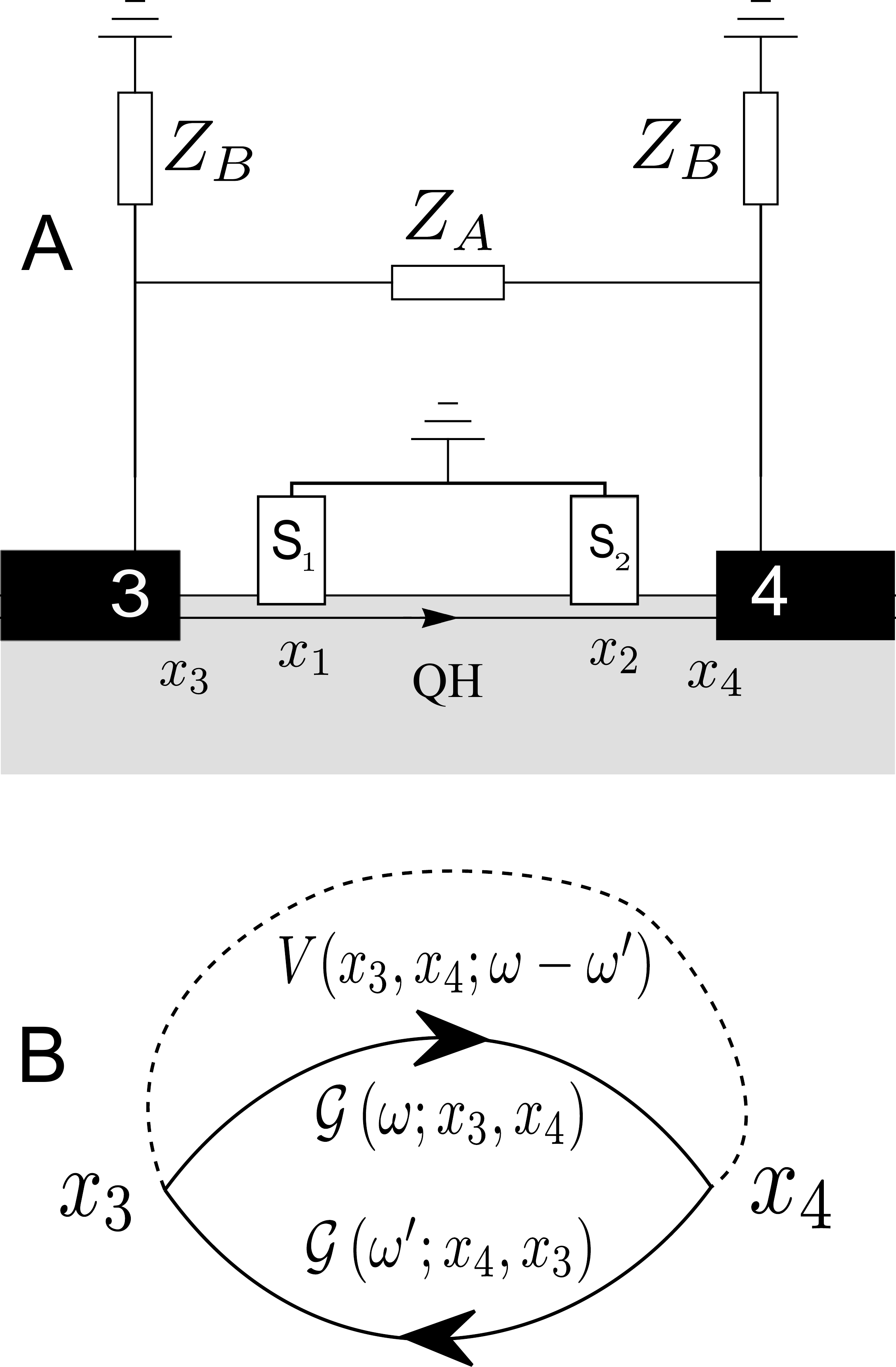}	
\caption{Interaction-induced supercurrent in a Quantum Hall edge channel. A. A chiral channel at the edge of the Quantum Hall bar (light-grey) is connected to two superconducting electrodes in the points $x_{1,2}$. Further away, the channel is covered with metallic electrodes $3,4$ that provide a non-local electron-electron interaction between the points $x_{3,4}$. B. The diagram for the relevant interaction correction to the total energy of the system, $\mathcal{G}$ being electron Green functions, $V$ being the interaction. The chirality of the channel requires that the diagram gives a non-zero contribution only if $\omega \omega'<0$.}
\label{fig:old}
\end{figure}
Here we introduce the microscopic model, shortly recite the results of Ref. \cite{HXL&YVN} and explain the mechanism of the interaction-induced supercurrent.

Let us consider a chiral channel at the edge of a Quantum Hall bar (Fig. \ref{fig:old}). We assume that the relevant energy scales are much smaller than the Landau level separation, thus the edge states can be described with a Hamiltonian encompassing electron field operators $\psi _{\sigma }(x)$ with a linear spectrum near Fermi level, $\sigma = \uparrow, \downarrow$ being the spin index 
\begin{equation}
H_0=-i v_F\sum _{\sigma } \int dx \, \psi _{\sigma }^{\dagger }(x)\partial _x \psi _{\sigma }(x) 
\end{equation}
In addition to this, we include the terms with the electrostatic potential $V(x)$ and the pairing potential $\Delta(x)$ induced to the channel in the vicinity of superconducting electrodes,
\begin{align}
H_1 =& \sum _{\sigma } \int dx \, V(x)\psi _{\sigma }^{\dagger }(x)\partial _x \psi _{\sigma }(x) \nonumber \\
+ \int dx& \,\left(\Delta^*(x) \psi_\uparrow(x) \psi_{\downarrow}(x) +\Delta(x) \psi^\dagger_\downarrow(x) \psi^\dagger_{\uparrow}(x)\right).
\end{align}
We have not considered the electrostatic potential in \cite{HXL&YVN}.

The resulting Matsubara Green function $\mathcal{G}(\omega; x,x')$ is a $2\times2$ matrix in Nambu space and satisfies 
\begin{align}
\label{eq:Green}
\left(-i \omega -i v_F \partial_x +\mathcal{H}(x)\right)\mathcal{G}\left(\omega ;x,x'\right)=-\delta (x-x') 
\\
\mathcal{H}(x) =V(x)\tau_z+ \Delta(x) \tau^+ + \Delta^*(x)\tau^-
\end{align}
where $\tau_z, \tau^{\pm} = (\tau_x \pm i \tau_y)/2$ are Pauli matrices in Nambu space. 

It is important to notice that the chirality of the channel is manifested in the form of Green functions as follows: $\mathcal{G}(\omega;x,x') = 0$ if $\omega>0, x>x'$ or $\omega<0, x<x'$. 
We compute supercurrent as a part of the energy that depends on the difference of the superconducting phases. For this correction, the Green functions should form a closed loop encompassing the coordinates of the superconducting terminals, $x_{1,2}$. The above property makes such loop impossible if the Green functions are at the same energy $\omega$. This forbids the supercurrent for non-interacting electrons. Such a loop is however possible if the frequencies of the Green functions making the loop are opposite in sign, this may be the case when they are the parts of the interaction correction (Fig. \ref{fig:old} b, the loop is formed if $\omega\omega' <0$)
The interaction that leads to supercurrent  must be a non-local one: since the loop encompasses $x_1,x_2$ the interaction line should connect the points $<x_1$ to those $>x_2$.

The general form of the interaction correction reads:
\begin{align}
(\Delta E)  = &  -2\int d\omega\ d\omega'\ dx\ dx' V\left(\omega -\omega';x
,x\right)			\nonumber \\
&	\text{Tr}\left[\mathcal{G}\left(\omega; x,x'\right)\hat{\tau }_z\mathcal{G}\left(\omega'; x',x\right)\hat{\tau }_z\right] 
\end{align}
Let us elaborate on the Green functions. Since those obey the first-order differential equation, its general solution for $\omega>0$
reads 

\begin{eqnarray}
v_F\mathcal{G}\left(\omega; x,x'\right)= e^{-\frac{\omega(x-x')}{v_F}}Pe^{-\frac{i}{v_F}\int_{x}^{x'}dz \mathcal{H}(z)}
\label{eq:GF_sol}
\end{eqnarray}
where $P$ indicates the position ordering of the exponent agrumens that are arranged from the left to the right in descending order of their coordinates.

The contributions of pairing and electrostatic potential to the position-ordered exponent are separated in space differently. The contributions of $\Delta$ come from the vicinity of each superconducting lead $i$, 
$\hat{U}_i = P\exp(-iv_F^{-1}\int dz (\Delta(z) \tau^+ + \Delta^*(z)\tau^-)$ where the integration interval covers the vicinity, and  are readily expressed in terms of the electron-hole conversion (Andreev) probability $p_i$ at this lead, 
\begin{equation}
U_i=\left(
\begin{array}{cc}
 \sqrt{1-p_i} & -i e^{i \phi _i}\sqrt{p_i} \\
 -i e^{-i \phi _i}\sqrt{p_i} & \sqrt{1-p_i} \\
\end{array}
\right).
\end{equation}
It is a unitary matrix that depends on the superconducting phase $\phi_i$ at this particular lead. The contribution of the electrostatic potential is accumulated on an interval $x_b>x_a$ and reads
\begin{equation}
\hat{K}_{ab} =\left(
\begin{array}{cc}
 e^{i \chi _{ab}} & 0 \\
 0 & e^{-i \chi_{ab}} \\
\end{array};\right)\; \chi_{ab} = -\int_{x_a}^{x_b} dx V(x)/v_F
\end{equation}
$\chi_{ab}$ being a dynamical phase\cite{QuantumTransport}accumulated over the interval. With this, for any interval $(x_3,x_4)$ that includes the superconducting electrodes the Green function reads
\begin{eqnarray}
\label{eq:Qform}
v_F\mathcal{G}\left(\omega; x_3,x_4\right)
&=&
-e^{-\frac{\omega}{v_F}(x_3-x_4)}
[\theta(\omega)\theta(x_3-x_4) \hat{Q}
\nonumber \\
&&-\theta(-\omega)\theta(x_4-x_3) \hat{Q}^{-1}
]
\end{eqnarray}
where a unitary matrix $\hat{Q}$ is the P-ordered exponent on this interval, 
\begin{equation}
\hat{Q} = \hat{K}_{31} \hat{U}_1 \hat{K}_{12} \hat{U}_2 \hat{K}_{24}
\end{equation}  
The energy correction contains a factor incorporating information about the Andreev reflection and superconducting phases,
\begin{equation}
A = {\rm Tr}[\hat{Q}\tau_z \hat{Q}^{-1}\tau_z]
\end{equation}
Since the matrices on the ends $\hat{K}_{31},\hat{K}_{24}$ commute with $\tau_z$, we can neglect those and reduce $Q$ to $Q' \equiv \hat{U}_1 \hat{K}_{12} \hat{U}_2$. With this,
\begin{align}
&A= 2(1-2 p_1)(1-2p_2) \nonumber \\ &-8\sqrt{p_1(1-p_1)p_2(1-p_2)} \cos(\phi_1-\phi_2 -2\chi_{21})
\end{align}

Let us specify the interaction to the model used in \cite{HXL&YVN}. To realize a non-local interaction that transfers the electric signals upstream, one embeds the QHE edge into an external electric circuit (Fig. \ref{fig:old}a).
To connect the edge to the circuit, we cover it with two metallic electrodes that are spread at $x<x_3$ and $x>x_4$ respectively, ($x_4-x_3\equiv \tilde{L}$) By a guage transform, the interaction can be reduced to the contact points and is expressed in terms of the cross-impedance between these electrodes (for the circuit in Fig. \ref{fig:old}a, $Z_{34}=Z_B^2/(Z_A+2Z_B)$)

\begin{equation}
V(\nu;x,x') = \frac{v_F^2}{2}\delta(x-x_3) \delta(x-x_4) \frac{Z_{34}(\nu)}{|\nu|}
\end{equation}

We specify to the model of the frequency-independent (at the scale $\simeq v_F/\tilde{L}$) impedance to arrive at 
\begin{align}
\Delta E = \frac{A R_{34}}{2}; \; R_{34}=\frac{e^2}{\pi^2} \frac{v_F}{\tilde{L}} Z_{34}. 
\end{align}
We compute the current by differentiating the energy with respect to the phase difference $\phi \equiv \phi_1 -\phi_2 $
\begin{align}
\label{eq:current}
&I(\phi)= 2 \partial_{\phi} (\Delta E) = 
 \\ &= -8  e R_{34} \sqrt{p_1 p_2 (1- p_1)(1- p_2)}\sin(\phi- 2\chi_{21})
\end{align}

This differs from the answer given in \cite{HXL&YVN} by the inclusion of the dynamical phase $\chi_{21}$ that effectively shifts the superconducting phase difference.  The dynamical phase is invariant with respect to  time reversal while the superconducting phase is not, so one may wonder why those two match each other. However, the time reversibility is essentially violated in QHE regime, and the chirality sets the relation between the phases. This leads to interesting and measurable consequences: the supercurrent between two electrodes can be modulated by a gate voltage applied to the channel to induce the dynamical phase. This effect of the gate voltage is rather local: it needs to be applied to the part of the channel between the superconducting electrodes. 

The scale of the current is $e$ times the inverse time of flight between the electrodes $v_F/\tilde {L}$ times a small factor that is the dimensionless impedance $Z_{34} e^2/\hbar$. A common estimation of for $Z_{34}$ is the vacuum impedance, this gives the small factor $\simeq 10^{-2}$.

Let us note that the current is a sinusoidal function of phase. In usual superconducting junctions, this occurs only in the limit of low transparency, and the corresponding process is identified as a single Cooper pair tunneling between the electrodes 1 and 2. Here the transparency is high since the channel is completely ballistic. Nevertheless, the underlying elementary process seems to be a single Cooper pair tunneling.

\section{Single constriction}
\label{sec:single}
\begin{figure}[h]
\includegraphics[width=.5\textwidth]{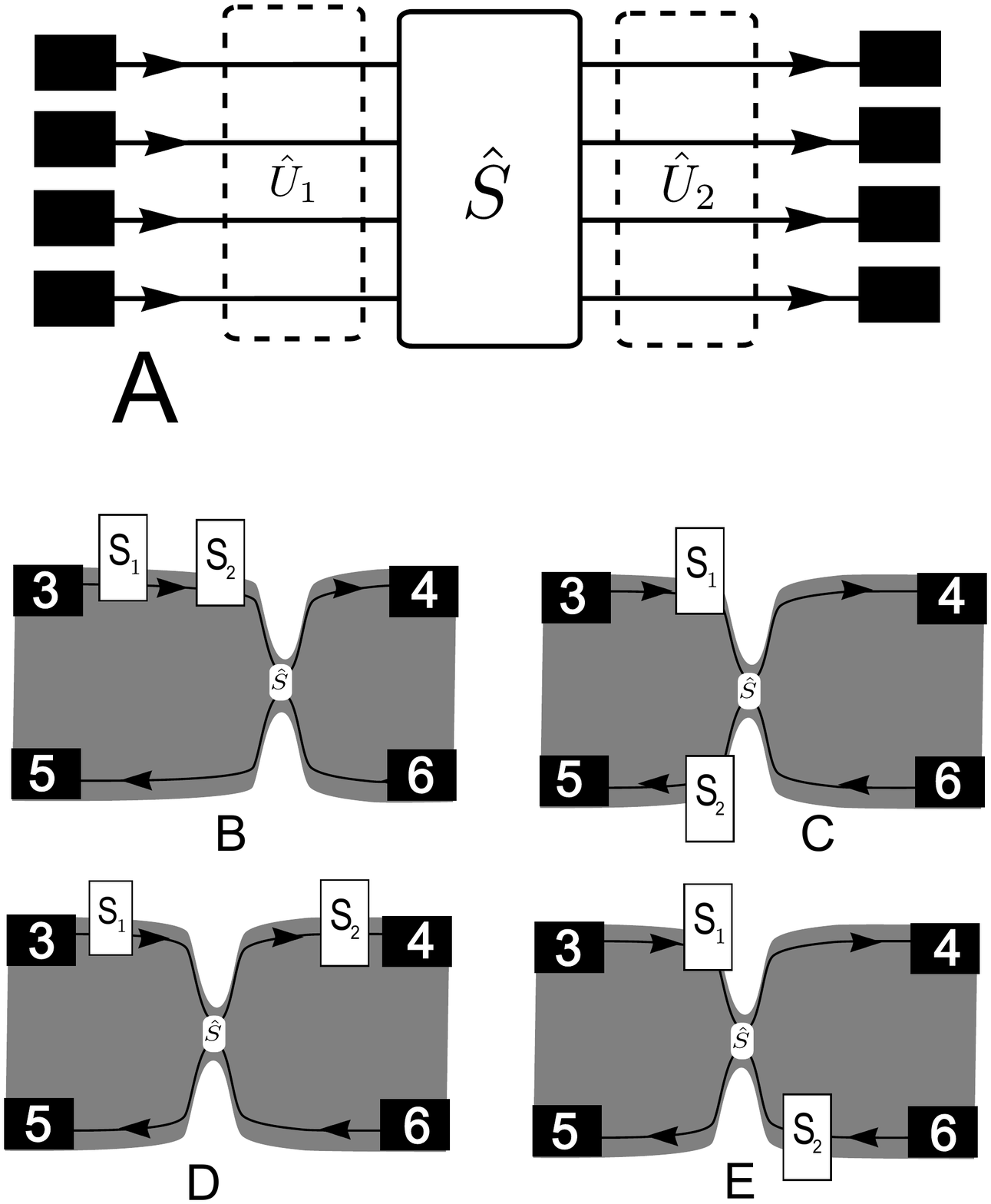}	
\caption{Single constriction setups. A. General framework for all possible single-constriction setups. We count the coordinates for each channel from the constriction in the direction of the channel propagation. In this way, we can describe Andreev and potential scattering on the same footing with a local potential that may mix $N$ propagating channels. This allows for an easy generalization of the single-channel approach. B-E. Various setups. The modulation of the constriction transmission changes the interaction coefficient in the setup B. and propagation probability in the setups D. and C. No interaction-induced supercurrent is found in setup E.  }
\label{fig:single}
\end{figure}

The simplest way to make a non-trivial QH setup is to make a constriction in a QH bar with the width that is comparable with the spread of the edge channel wavefunctions (Figs. \ref{fig:single}) B-E). There is a scattering of the electron waves at the constriction: upon passing the constriction, an incoming electron will either stay at the same edge with probability $T$ or be reflected to the opposite edge with probability $1-T$. In all setups, we implement the external circuit non-local interaction: the beginning and end of each channel is covered by a metal electrode included in the circuit (not shown in the Figure). 

A common specific feature of all single-constriction setups is that the electron trajectories do not form any closed loops whatever the scattering they experience. This is why all such setups can be treated in the same manner. We do this by counting the coordinates for each channel separately in the direction of propagation, starting form a point in the constriction. In this way, we can reduce all setups to a single general model depicted in Fig. \ref{fig:single} a. There, we have $N$ chiral channels subject to local pairing and electrostatic potential, these potentials being $N\times N$ matrices in the channel space. The Green function is also a matrix in channel space satisfying 
\begin{align}
\label{eq:Greenchannels}
\left(-i \omega -i \check{v}_F \partial_x +\mathcal{H}(x)\right)\mathcal{G}\left(\omega ;x,x'\right)=-\check{1}\delta (x-x') 
\\
\mathcal{H}(x) =\check{V}(x)\tau_z+ \check{\Delta}(x) \tau^+ + \check{\Delta}^*(x)\tau^-
\end{align}
where "check" denotes the matrix structure in the channel space.
We note that $v_F$ also has this structure since the velocity may depend on the channel. 
Apart from this extra structure, the Eq. \ref{eq:Greenchannels} is a complete analogue of Eq. \ref{eq:Green} and can be solved with a position-ordered exponent.

To simplify further, we note that pairing potential is diagonal in channels, either before or after the constriction, and the non-diagonal potential is localized on the constriction (Fig. \ref{fig:single} a). With this, the Green function can be represented in a form analogous to Eq. \ref{eq:Qform}, $a,b$ being the channel indices, $x,x'$ are beyond the scattering region,
\begin{eqnarray}
\label{eq:Qforme}
\sqrt{v^a_Fv^b_F}\mathcal{G}^{ab}\left(\omega; x,x'\right)
&=&
-e^{-\frac{|\omega|}{v_F}|x-x'|)}
[\theta(\omega)\theta(x-x') \hat{Q}^{ab}
\nonumber \\
&&-\theta(-\omega)\theta(x'-x) (\hat{Q}^{-1})^{ab}
]
\end{eqnarray}

The unitary matrix $\hat{Q}$ is composed of the matrices of the superconducting electrodes before and after the constriction, and the matrix $\hat{S}$ that describes the scattering at the constriction,
\begin{equation}
\hat{Q} = \hat{U}_1 \hat{S} \hat{U}_2.
\end{equation}
For two channels,
\begin{equation} 
\hat{S}=\check{s}\frac{1+\tau_z}{2}+\check{s}^\dagger \frac{1-\tau_z}{2};\; \check{s} \equiv \left(
\begin{array}{cc}
 t & r \\
 -r' & t' \\
\end{array};\right)
\end{equation}
$t,t'$ and $r,r'$ being the transmission and reflection amplitudes at the constriction.  
To compute the interaction correction, we employ the external circuit non-local interaction model. In general, we have the contributions from each pair of the electrodes at the beginning and at the end of the channel, those are weighted with the corresponding cross-impedances. 

We are ready to derive the answers for the specific setups. Let us start with one shown in Fig. \ref{fig:single} B. Here, both superconducting electrodes are connected to the same channel upstream from the constriction. Naively, one would regard the superconducting current as a local quantity determined by the electrodes and the space between those. However, this is not true in view of the non-local character of the interaction. The setup provides a good and practical illustration for this. 
Similar to Eq. \ref{eq:current}, the current is given by 
$I=-8  e R_{B} \sqrt{p_1 p_2 (1- p_1)(1- p_2)}\sin(\phi- 2\chi_{21})$
with 
\begin{equation}
R_B = \frac{e^2}{\pi^2} \left(T \frac{v_F}{L_3+L_4} Z_{34} +(1-T)\frac{v_F}{L_3+L_5} Z_{35}\right)
\end{equation}
Here and further in the text, $L_i$ is the distance from the constriction to the metallic electrode $i$.
We see that the interaction coefficient $R$ does depend on the transmission of a distant constriction switching between two values corresponding to completely open and closed constriction. A gate that modulates this transmission will modulate this interaction coefficient and current without changing the external circuit. This would be a convenient experimental proof of non-locality. Similar result is obtained if both electrodes connect the same channel downstream the constriction.

If one electrode is upstream from the constriction, and another one is downstream, (Fig. \ref{fig:single} C,D), the modulation of the transmission modifies the probability to go from one to another, rather than the interaction. The current is given by $I=-8  e R_{C,D} \sqrt{p_1 p_2 (1- p_1)(1- p_2)}\sin(\phi- 2\chi_{21})$ 
with 
\begin{equation}
R_C = \frac{e^2}{\pi^2} (1-T) \frac{v_F}{L_3+L_5} Z_{35};\; R_D = \frac{e^2}{\pi^2} T \frac{v_F}{L_3+L_4} Z_{34}
\end{equation}
The dynamical phase $\chi_{21}$ is accumulated along a path passing the constriction and eventually incorporates the phase of either transmission or reflection amplitude. Interestingly, $R_B = R_C +R_D$, this can be used for the experimental identification of the effect.

For a setup where the superconducting electrodes are either upstream or downstream from the constriction but contact different channels  (Fig. \ref{fig:single} E), we find no interaction-induced supercurrent.
This is related to the fact that one cannot make a closed loop of Green functions encompassing both electrodes. Beenakker \cite{Beenakker} has proposed to measure current noise correlations in the setup. Luckily, those would not be obscured by the supercurrent.

\section{Multiple superconducting terminals}
\label{sec:multi}
\begin{figure}[h]
\includegraphics[width=.5\textwidth]{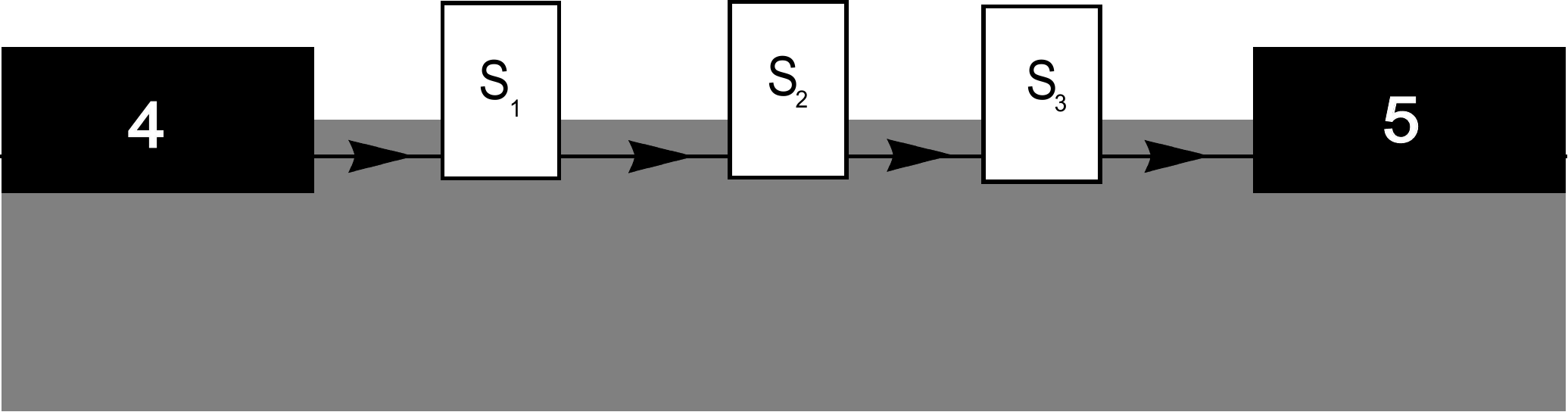}	
\caption{Multiple superconducting electrodes contacting the same channel, $N=3$. }
\label{fig:multisuper}
\end{figure}

In view of a significant experimental and theoretical interest to multi-terminal superconducting nanostructures, we consider here multiple superconducting electrodes connected to the same channel. The approach outlined in the previous Section suits for multiple superconducting electrodes as well. Here, we concentrate on a simple but general situation when $N$ superconducting electrodes are in contact with the same channel  (Fig. \ref{fig:multisuper} gives the setup for $N=3$).

The Green function between the edges of metallic electrodes 
is given by Eq. \ref{eq:Qform} with $Q$ encompassing all matrices $\hat{U}_i$, $i=1..N$, of the superconducting electrodes and the matrices $\hat{K}_{i,i+1}$ responsible for the accumulation of dynamical phase between the electrodes,
\begin{equation}
\label{eq:QAllU}
\hat{Q} = \prod_{i=1^{N-1}} \hat{U}_i  \hat{K}_{i,i+1} \cdot \hat{U}_N
\end{equation}
Here we skip $\hat{K}$ matrices before and after the superconducting electrodes since they do not affect the answer for the current.

The energy correction is given by $\Delta E = A R/2$, where the interaction coefficient is fixed to $\frac{e^2}{\pi^2} \frac{v_F}{\tilde{L}} Z_{45}$ while 
\begin{equation} 
A = 2(1-2|Q_{eh}|^2)
\end{equation}
incorporates all the information about the Andreev probabilities and superconducting phases.

Actually, $Q_{eh}$ is the amplitude of Andreev conversion of an electron to a hole while passing the setup, and $|Q_{eh}|^2$ is the conversion probability. It is instructive to regard it as a sum of partial Andreev amplitudes corresponding to different sequences of conversion or passing at the electrodes. For instance, there are partial amplitudes where the electron is converted at one of the electrodes passing all other. Another set of the partial amplitudes corresponds to the case  when the election is converted to the hole at the first electrode, the hole is converted back to the electron at the second, and finally back to hole at the third one, while passing all others. Each partial amplitude, in agreement with Eq. \ref{eq:QAllU}, is a product of amplitudes from all electrodes and spaces in between those. Let us give an example of such analysis for $N=3$ and derive the expression for $A$.

The amplitude of the process where the conversion occurs at the electrode $1$ reads
\begin{equation}
\mathcal{A}_1 = -i \sqrt{p_1} \sqrt{1-p_2} \sqrt{1-p_3} e^{i\phi_1} e^{-i\chi_{12}} e^{-i\chi_{23}},
\end{equation}
similar contributions for electrodes $2$, $3$ are obtained by index exchange and change of signs of the dynamical phase:
\begin{align}
\mathcal{A}_2 = -i \sqrt{p_2} \sqrt{1-p_1} \sqrt{1-p_3} e^{i\phi_2} e^{i\chi_{12}} e^{-i\chi_{23}},\\
\mathcal{A}_3 = -i \sqrt{p_3} \sqrt{1-p_2} \sqrt{1-p_1} e^{\phi_3} e^{\chi_{12}} e^{\chi_{23}},
\end{align}
and there is a contribution that corresponds to the conversion at each electrode,
\begin{equation}
\mathcal{A}_{123} = (-i)^3 \sqrt{p_1}\sqrt{p_2}\sqrt{p_3} e^{i\phi_1} e^{-i\chi_{12}} e^{-i\phi_2} e^{i\chi_{23}} e^{i\phi_3} . 
\end{equation}
Let us for convenience shift the phases $\phi_{1,3}$ with the corresponding dynamical phases, $\phi_1 \to \phi_1 - 2\chi_{12}$, $\phi_3 \to \phi_3 + 2 \chi_{23}$. With this, we express the conversion probability as
\begin{align}
&|Q_{eh}|^2 = p_2(1-p_1)(1-p_3)+p_1(1-p_2)(1-p_3) \nonumber \\
&+ p_3(1-p_2)(1-p_1) + p_1 p_2 p_3 \label{term0}\\
&+2 \sqrt{p_1 (1-p_1) p_2(1-p_2)}\cos(\phi_1-\phi_2) \label{term12}\\
&+2 \sqrt{p_2 (1-p_2) p_3(1-p_3)} \cos(\phi_2-\phi_3) \label{term23}\\
&+2 \sqrt{p_1 (1-p_1) p_3(1-p_3)}(1-p_2)
\cos(\phi_1-\phi_3) \label{term13}\\
&-2 \sqrt{p_1 (1-p_1) p_3(1-p_3)}p_2 
\cos(2\phi_2-\phi_3-\phi_1) \label{termnew}
\end{align}

Here, the term (\ref{term0}) compises the squares of the partial amplitudes. It does not depend on phases and therefore does not contribute to the current. The term (\ref{term12}) comes about the interference of the amplitudes in pairs $\mathcal{A}_1$, $\mathcal{A}_2$ and $\mathcal{A}_3$, $\mathcal{A}_{123}$. Somewhat surprisingly, it corresponds to the currents between the electrodes 1 and 2 as if the third electrode was not at all present. The same applies to the term (\ref{term23}): it corresponds to the current between the electrodes 2 and 3 as if no electrode 1 is present and arises from the interference of the amplitudes in pairs $\mathcal{A}_3$, $\mathcal{A}_2$ and $\mathcal{A}_1$, $\mathcal{A}_{123}$. The term (\ref{term13}) describes the current between 1 and 3 only, although its amplitude is reduced by Andreev conversion at the electrode 2, and manifests interference between $\mathcal{A}_1$ and $\mathcal{A}_3$. All these terms lead to the currents as if there were tunnel junctions connecting the corresponding electrodes and manifest a single Cooper pair tunneling between the electrodes. 
The last term (\ref{termnew}) is of different nature. It manifests a more interesting process of two Cooper pair tunneling: the Cooper pairs  from 1 and 3 simultaneously entering the electrode 2, or resersely, two Cooper pairs from 2 getting to 1 and 3, that cannot be described with elementary tunnel junctions. For bigger number of electrodes, more complex processes involving more electrodes and Cooper pairs, are manifested.

Finally, the currents read (we count the phases from the electrode 2, $\phi_2=0$)
\begin{align}
I_1 &= I^0_1 \sin \phi_1 + I_{13} \sin (\phi_1-\phi_3) +I_{i} \sin (\phi_1+\phi_3), \\
I_3 &= I^0_3 \sin \phi_1 + I_{13} \sin (\phi_3-\phi_1) +I_{i} \sin (\phi_1+\phi_3), 
\end{align}
$I_2 = - I_3 - I_1$, where $I_{i}^0 = - 8 e R_{45} \sqrt{p_i p_2 (1- p_i)(1- p_2)}$, $I_{13} = - 8 e R_{45} \sqrt{p_1 p_3 (1- p_1)(1- p_3)}(1- p_2)$, $I_i = - 8 e R_{45} \sqrt{p_1 p_3 (1- p_1)(1- p_3)}p_2$. The last terms $\propto I_i$ are due to the interesting process. The currents are shifted sinusoidal functions of any phase.

\section{An example of a complex setup}
\label{sec:complex}
\begin{figure}[h]
\includegraphics[width=.5\textwidth]{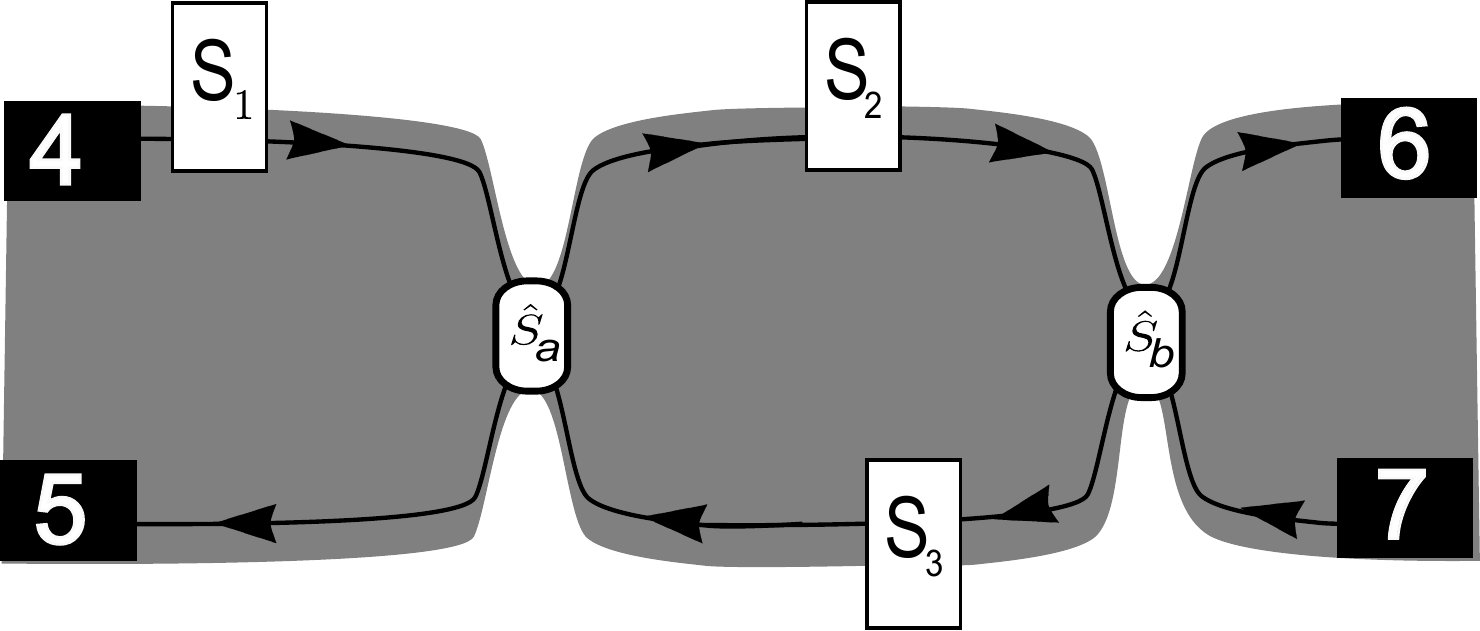}	
\caption{An example of a more complex setup: A Hall bar with two constrictions. This provides a possibility of  electron trajectories that loop over the ring between the constrictions $a$ and $b$. This leads to non-interacting current between the electrodes $2$ and $3$. The supercurrent to $1$ is interaction-induced and is evaluated in this Section. }
\label{fig:twoconstrictions}
\end{figure}

It is not difficult to form two constrictions in a Hall bar (Fig. \ref{fig:twoconstrictions}) This provides an example of more complex setup that cannot be understood with the approach of the previous sections. The reason for this is a possibility of looping electron trajectories that provide multiple Andreev conversions from the same electrodes multiple scatterings at the same constriction. In the setup under investigation, the loops occur in the ring between the constrictions $a$ and $b$. 

It has to be noted that looping trajectories lead to a non-interacting current, in this case, between the electrodes 2 and 3. The magnitude of this current can be estimated as $e v_F/L_c$ and is typically much bigger than the expected interaction-induced current. The precise expression can be derived from the phase-dependent contribution to the ground state energy that reads 
\begin{equation} 
\Delta E =  - \frac{v_F}{L_c} \sum_{pm} {\rm Li}_2(\sqrt{R_aR_b} e^{\pm i \lambda})
\end{equation}
Here, $e^{\pm i \lambda}$ are the eigenvalues of the matrix $\hat{Q}_c = \hat{U}_2 \hat{K}_{23} \hat{U}_3 \hat{K}_32$, the matrices $\hat{U}_{2,3}$ represent Andreev conversion at the corresponding electrodes while $\hat{K}_{23,32}$ represent the accumulation of dynamical phases on paths $2\to3$, $3\to2$ and include the phases of the reflection amplitudes $r_a, r'_b$, respectively. More explicitly
\begin{align}
&\cos\lambda = Q_0 = \cos(\chi_{23}+\chi_{32}) \sqrt{(1-p_2)(1-p_3)} \nonumber \\
&+ \sqrt{p_2p_3}\cos(\phi_2-\phi_3 +\chi_{32} -\chi_{23}) \label{eq:Q0}
\end{align}

However, the supercurrent from the electrode $1$ can only be due to a non-local interaction. Let us compute the contribution proportional to $Z_{45}$, all other contributions can be evaluated in the same manner. We start with evaluation of $\mathcal{G}(\omega, x_4,x_5)$ at $\omega>0$. It is determined by electron-hole propagation between these points and is a sum of partial propagation amplitudes with different number of loops in the ring between the constrictions $a$ and $b$. 
The contribution with no loops encompasses the propagation along the paths $4\to a \to 5$ and reads
\begin{equation}
\mathcal{G}^{(0)} = e^{-\frac{\omega(L_4+L_5)}{v_F}}\hat{\bar{U}}_1\hat{r}_a 
\end{equation}
Here, to shorten the notations, we introduce $\hat{\bar{U}}$ that incorporates the adjacent $\hat{K}$ (for instance, $\hat{\bar{U}}_1 = \hat{K}_{51} \hat{U}_1 \hat{K}_{1a}$) skip the irrelevant $\hat{K}$ at the end of the path.
The contribution with one loop, in addition to this, encompasses the path $a \to 2 \to b \to 3 \to a$,
\begin{equation}
\mathcal{G}^{(1)} = e^{-\frac{\omega(L_4+L_5)}{v_F}}\hat{\bar{U}}_1 \hat{t}_a e^{-\frac{\omega L_c}{v_F}} \hat{\bar{U}}_2 \hat{r}_b \hat{\bar{U}}_2
\end{equation}
The contributions with higher loop numbers form a geometric series where each term being multiplied with $\hat{r}'_a  \hat{\bar{U}}_2 \hat{r}_b \hat{\bar{U}}_3 \exp(-\omega L_c/v_F) \equiv \hat{Q}_c \sqrt{R_a R_b} \exp(-\omega L_c/v_F)$. This sums up to
\begin{align}
\mathcal{G} &= e^{-\frac{\omega(L_4+L_5)}{v_F}}\hat{\bar{U}}_1 \left(\hat{r}_a+( \hat{t}_a(\hat{r}'_a)^{-1} \hat{M} \hat{t}'_a \right);\\
\hat{M} &\equiv \frac{\hat{Q}_c \sqrt{R_a R_b} e^{-\frac{\omega L_c}{v_F}}}{1 -\hat{Q}_c \sqrt{R_a R_b} e^{-\frac{\omega L_c}{v_F}}}.
\end{align}
We concate this with another Green function at $\omega'$ and integrate over $\omega,\omega'$. This intergation is more involved than in the previous cases since the propagation involves the paths of different lenghtes, $L = L_4+L_5 + n L_c$, $n$ being the number of the loops made by a trajectory. The answer involves many different combinations of dynamical phases. To simplify, we 
shift $\phi_2 \to \phi_2 +\mu +\nu$, $\phi_3 \to \phi_3 + \mu - \nu$, $\mu \equiv {\rm arg}(r'_a) + \chi_{a2}$, $\nu \equiv \chi_{2b} +{\rm arg}(r_b)+ \chi_{b3}$, and introduce
\begin{align}
\chi_A &= 2\chi_{1a} +{\rm arg}(t_a) - {\rm arg}(r'_a) -{\rm arg}(t'_a)  -\chi_{3a}\\
\chi_B &= 2\chi_{1a} +2{\rm arg}(t_a) - 2{\rm arg}(r'_a)+\nu+\mu\\
\chi_C &= 2\chi_{1a} +2{\rm arg}(t_a) - 2{\rm arg}(r'_a)-\chi_{3a}
\end{align}
With this,
\begin{align}
&\Delta E = - 4\sqrt{p_1(1-p_1)} T_a \frac{e^2}{\pi^2} \frac{v_F}{L_4+L_5} Z_{34} [ \\
&\sqrt{R_aR_b} {\cal F}_A(\cos(\chi_A -\phi_1 +\phi_2) \sqrt{p_2(1-p_3)} \nonumber\\
& + \cos(\chi_A -\phi_1 +\phi_3) \sqrt{p_3(1-p_2)}) \label{sortA}\\
&+2T_a R_b {\cal F}_B (-\cos(\chi_B -\phi_1 +\phi_2) \sqrt{p_2(1-p_2)} (1-p_3) \nonumber\\
& - \cos(\chi_B -\phi_1 +\phi_3) \sqrt{p_3(1-p_3)} (1-p_2) \nonumber \\
& + \cos(\chi_B -\phi_1 -\phi_3 +2 \phi_2) \sqrt{p_3(1-p_3)} p_2)\label{sortB}\\
&+T_a R^{3/2}_b R^{1/2}_{a} {\cal F}_C(\cos(\chi_C -\phi_1 +\phi_2) \sqrt{p_2(1-p_3)} \nonumber \\
&+\cos(\chi_C -\phi_1 +\phi_3) \sqrt{p_3(1-p_2)})] \label{sortC}
\end{align}

Here, the dimensionless coefficients ${\cal F}_{A,B,C}$ come about the frequency integration. They depend on the ratio of paths $c \equiv L_c/(L_4+L_5)$, the reflection coefficient $\sqrt{R_AR_B}$ in the ring and incorporate information about the Andreev conversion, superconducting and dynamical phases in the ring by a single parameter $Q_0$ defined by Eq. \ref{eq:Q0}

The coefficients are expressed in integral form as
\begin{align}
{\cal F}_A = \int_0^{\infty}\frac{dx \ dx'}{x+x'} &e^{-(x+x')} \left(e^{-cx} D(x) \right.\nonumber \\
&\left.+ e^{-cx'} D(x')\right); 
\end{align}
\begin{equation}
{\cal F}_B = \int_0^{\infty}\frac{dx \ dx'}{x+x'} e^{-(x+x')(1+c)} D(x)D(x');\end{equation}
\begin{align}
&{\cal F}_B = - \int_0^{\infty}\frac{dx \ dx'}{x+x'} e^{-(x+x')(1+c)} \left(e^{-cx'} \right.\nonumber \\
&\left.+ e^{-cx}\right) D(x)D(x'); 
\end{align}
\begin{align}
&D^{-1}(x) = 1 - 2\sqrt{R_aR_b} Q_0 e^{-cx} + R_a R_b e^{-2cx}
\end{align}
They approach constant limits at $c\to 0$ and scale as $1/c$ at $c \to \infty$, this signifies that in the limit of large ring circumferences the energy scale is determined by $L_c$. 

Let us consider and interpret the terms in the phase-dependent energy correction. We see that the overall expression is proportional to $T_a$, since the electrons should get to the ring to feel other superconducting ring. The terms (\ref{sortA}) proportional to ${\cal F}_A$ come about the interference of paths that do and do not visit the ring, this is seen from square-root dependence on the reflection coefficients. We have not encountered this situation in the previous sections, since there any relevant path passes all the electrodes. We see this in different dependence of the coefficients on Andreev conversion probabilities, for instance, $\sqrt{p_2(1-p_3)}$ misses the factor $\sqrt{1-p_2}$ present in the previous expressions. The phase dependence of the terms can be still interpreted in terms of single Cooper pair tunneling between either 1 and 2, or 1 and 3. 

The terms (\ref{sortB}) proportional to $\cal{F}_B$ arise from the interference of various trajectories that visit the ring and experience Andreev conversion when going from 4 to 5. Their structure is similar to that studied in the previous section. There is a term that manifests a process whereby two Cooper pairs from 1 and 3 enter the electrode 2. It has to be present, since in the limit $R_b \to 1$, $R_a \to 0$ we return to the single-channel setup considered in Section \ref{sec:multi} where the two Cooper pair tunneling has been identified.
The terms (\ref{sortC})  proportional to $\cal{F}_C$ result from the interference of the trajectories that pass the ring with and without Andreev conversion in the ring. This is why the dependence of the coefficients on the conversion probabilities is identical to that of (\ref{sortA}). The presence of looping trajectories and various paths leads to the fact that the similar terms pick up different dynamical phases that cannot be compensated with the shifts of the superconducting phases as in the previous examples.

\section{Conclusions}
\label{sec:conclusions} 
In conclusion, we have extended the previous study of interaction-induced supercurrents in a single Quantum Hall edge channels to experimentally relevant and widely used Quantum Hall setups with scattering between the edge channels. We resticted ourselves to a simple but relevant interaction model where the informational flow in the direction opposite to that of the electron propagation is provided by an external circuit.

For a single constriction in a Hall bar, the considerations are simple and can be done in very general form. We have considered specific setups that manifest the non-local nature of the interaction-induced effect whereby a supercurrent can be modulated by changing the transmission coefficient of a distant constriction. We have considered a multi-terminal superconducting system where the electrodes are connected to a single edge channel, understood the supercurrent in terms of interference of Andreev conversion processes and have identified a process that can be regarded as two Cooper pair tunneling. 

We have considered a more complex exemplary setup  that involves two constrictions and thereby gives a possibility of looping trajectories. This gives rise to interplay of non-interacting and interaction-induced currents and significantly complicates the situation. We demonstrate the evaluation of the phase-dependent energy correction in this complex setup and interpret the result in terms of various interference processes. 

The results presented facilitate the experimental observation of interaction-induced supercurrent and contribute to the active field of superconductor-QHE nanostructures.

This research was supported by the Netherlands Organization
for Scientific Research (NWO/OCW).

\bibliography{paper}

\end{document}